\documentclass{svjour3}                     % onecolumn (standard format)
\smartqed  % flush right qed marks, e.g. at end of proof
\usepackage{graphicx}
\usepackage{fix-cm}
\usepackage{bm}
\journalname{Journal of Low Temperature Physics}
\begin{document}

\title{\boldmath Vortex core contribution to textural energy in $^3$He-B below
  $0.4\,T_{\rm c}$%\thanks{Grants or other notes
}

\titlerunning{Vortex core contribution to textural energy}

\author{V.B. Eltsov \and
        R. de Graaf \and
        M. Krusius \and
        D.E. Zmeev
}


\institute{V.B. Eltsov, R. de Graaf, M. Krusius \at
              Low Temperature Laboratory, Aalto University, Espoo, Finland \\
              Tel.: +358-9-47022973\\
              Fax: +358-9-47022969\\
              \email{ve@boojum.hut.fi}           %  \\
           \and
           D.E. Zmeev \at
             School of Science and Astronomy, 
             The  University of Manchester, UK
}

\date{Received: date / Accepted: date}

\maketitle

\begin{abstract}
  Vortex lines affect the spatial order-parameter distribution in
  superfluid $^3$He-B owing to superflow circulating around vortex cores
  and due to the interaction of the order parameter in the core and in the bulk
  as a result of superfluid coherence over the whole volume. The step-like
  change of the latter contribution at $0.6T_{\rm c}$ (at a pressure of
  29\,bar) signifies the transition from axisymmetric cores at higher
  temperatures to broken-symmetry cores at lower temperatures. We extended
  earlier measurements of the core contribution to temperatures below
  $0.2T_{\rm c}$, in particular searching for a possible new core
  transition to lower symmetries. As a measuring tool we track the energy levels of
  magnon condensate states in a trap formed by the order-parameter
  texture.
\end{abstract}

\section{Introduction}
\label{intro}

In superfluid $^3$He Cooper pairing in states with spin $S=1$ and orbital
momentum $L=1$ leads to multi-component order parameters and a variety of
topological defects, including quantized vortex lines of many different structures
\cite{vortrev}. The simplest vortex structure in any superfluid is a $2\pi$
phase
vortex with a hard singular core, i.e. with a core of approximately coherence
length $\xi$ in radius, where all order parameter components go to zero at
the axis. Such vortices exist in superfluid $^4$He and in superconductors
with conventional s-wave pairing. In systems with multi-component order
parameters, like $^3$He, such a simple possibility is never realized.
In vortices with the lowest energy the order parameter in the core may
deviate from that in the bulk, but remains finite throughout the whole
core. Such vortices with a hard, but non-singular core are realized in
the B phase of superfluid $^3$He.

\begin{figure}
\centerline{\includegraphics[width=0.8\linewidth]{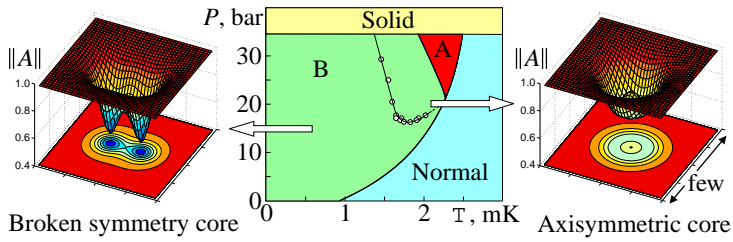}}
\caption{(color online) Vortex core transition in superfluid $^3$He-B. The
  phase diagram in the middle shows, besides the phases of $^3$He, the
  transition line between the axisymmetric core in the high pressure -- high
  temperature corner and the non-axisymmetric core in the rest of the
  $^3$He-B parameter
  space \cite{coretransdiagr}. The plots on the left and right show the
  calculated core amplitudes $||A||$ of the order parameter, normalized to the value
  in the bulk \cite{corecalc}.}
\label{diagr} 
\end{figure}

In fact, multiple choices for the order parameter structure in the
non-singular core are possible. $^3$He-B was the first system where
a transition between two different core structures was found experimentally
\cite{coretrans}. As seen in the phase diagram in Fig.~\ref{diagr}, vortices,
which are stable in $^3$He-B at temperatures above about $0.6\,T_{\rm c}$
and pressures above about 15\,bar, possess axisymmetric cores, like
vortices in most other known systems. In vortices, which are stable at
lower pressures or temperatures, the axial symmetry is broken. Each such
vortex can be viewed as a bound pair of two half-quantum vortices.

Superfluid coherence is preserved
across the whole volume when vortices with non-singular cores are present. The rigidity of
the order parameter (supported by the gradient energy) leads to a smooth
variation of the order
parameter across the sample, which is called a texture. In bulk $^3$He-B in a magnetic field
$\mathbf{H}$ the order parameter has the
form
\begin{equation}
A_{\mu j} = \Delta_{\mu \nu} R_{\nu j}(\hat\mathbf{n},\theta) e^{i\varphi},
\label{opar}
\end{equation}
where $\varphi$ is the superfluid phase, $\Delta_{\mu\nu}(\mathbf{H})$ is
the uniaxial gap matrix in a magnetically distorted B phase and $R_{\nu
  j}(\hat\mathbf{n},\theta)$ is the matrix of rotation around the axis
$\hat\mathbf{n}$ by the angle $\theta$ which characterizes the broken relative
spin-orbit symmetry. The spin-orbit interaction fixes $\theta$ at the value
$\theta = \arccos(-1/4)$. Thus the texture is essentially formed by the
spatial variation of the unit vector $\hat\mathbf{n}$. Alternatively, the
texture of the orbital anisotropy axis $\hat\mathbf{l} = (\mathbf{H}/H)
R(\hat\mathbf{n},\theta)$, which is defined in the presence of the magnetic
field, can be considered.

Various energies affect the orientations of the $\hat\mathbf{l}$ and
$\hat\mathbf{n}$ unit vectors in the texture \cite{Thuneberg}. The magnetic
anisotropy energy prefers $\hat\mathbf{l} \parallel \mathbf{H}$. The walls of
the container orient $\hat\mathbf{l}$ perpendicular to
themselves. A counterflow $\mathbf{v}_{\rm s} - \mathbf{v}_{\rm n}$ between
superfluid and normal velocities
$\mathbf{v}_{\rm s}$  and $\mathbf{v}_{\rm n}$ tends to pull $\hat\mathbf{l}$ along the flow. Here we are most interested
in the contribution of vortex lines to the textural energy. It can be
written in the form
\begin{equation}
F_{\rm vort} = \frac25 aH^2 \frac\lambda\Omega \int d^3r 
      \frac{(\bm{\omega}_{\rm v} \cdot \hat\mathbf{l})^2}{\omega_{\rm v}}~,
\label{fvort}
\end{equation}
where $a$ is the magnetic anisotropy parameter, $\bm{\omega}_{\rm v} =
\frac12 \langle \nabla \times {\bf v}_{\rm s}\rangle$ is the vorticity,
$\Omega$ is the rotation velocity, and $\lambda$ is the dimensionless parameter
characterizing the vortex contribution to the textural energy. For the equilibrium array of vortex lines $F_{\rm
  vort}$ is proportional to the number of vortices and hence $\lambda \propto
\Omega$. Later we will use $\lambda/\Omega$ as an intrinsic 
velocity-independent parameter.

The vortex effect on the texture has two contributions: $\lambda =
\lambda_{\rm f} + \lambda_{\rm c}$. The first contribution $\lambda_{\rm
  f}$ comes from the quantized superflow circulating around vortex cores and is
independent of the core structure. The other contribution $\lambda_{\rm c}$
comes directly from the cores since the order parameter has to change
smoothly between its bulk value and the value in the core. The latter
contribution is of special interest since it depends on the vortex core
structure. The step-like change in $\lambda_{\rm c}$ was in fact the
first experimental signature of the transition in the vortex core \cite{coretrans}.

The cores of $^3$He-B vortices have broken parity, which leads to
a gyromagnetic effect: The value of $\lambda_{\rm c}$ depends on the
orientation of the vortices with respect to the magnetic field direction:
$\lambda_{\rm c} = \tilde\lambda_{\rm c} \pm \kappa/H$, where $\kappa > 0$
is the gyromagnetic parameter, the plus sign refers to vortices with their circulation oriented
along the magnetic field and the minus sign is for the opposite orientation.
Both $\tilde\lambda_{\rm c}$ and $\kappa$ change at the core transition \cite{gyro}.

In this work the previous measurements of $\lambda$ at $T>0.5\,T_{\rm c}$ \cite{HakonenJLTP},
have been extended to below $0.2\,T_{\rm c}$. One of the goals was
to check for the possible existence of other transitions in the vortex
core structure. In Sec.~\ref{sec:meas} we explain the general principles of
probing the texture using spin-wave resonances. In Sec.~\ref{sec:nmr} we
describe a novel precessing state, related to Bose-Einstein condensation of
magnons, which is used to determine the positions of the resonances at low
temperatures. In Sec.~\ref{sec:res} we present our results for the textural
energy parameters $\lambda$ and $\kappa$. Finally in Sec.~\ref{sec:dyn} we
demonstrate how the NMR techniques developed for the measurements of $\lambda$
can be used to probe vortex dynamics at the lowest temperatures.

\section{Principles of the measurement}
\label{sec:meas}

\begin{figure}
\centerline{\includegraphics[width=0.8\linewidth]{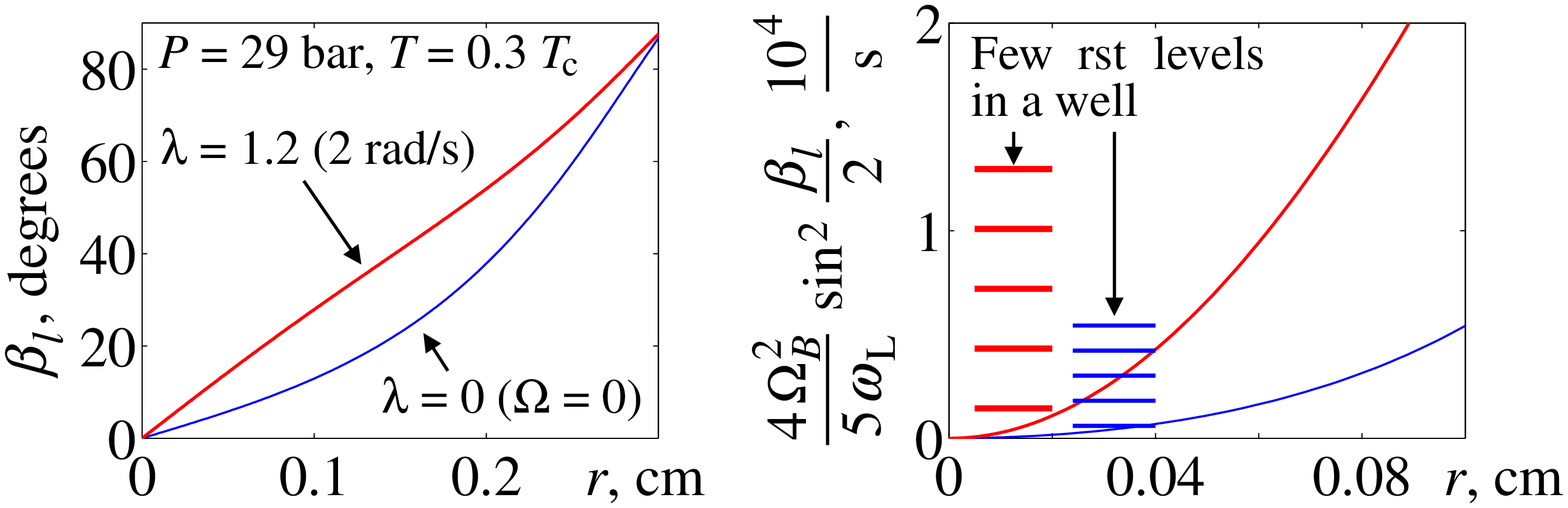}}
\caption{(color online) \textit{(Left)} Order parameter texture for two
  values of the vortex textural energy parameter $\lambda$, which
  correspond to the equilibrium state in rotation at $\Omega=2\,$rad/s and
  at rest. (Results of
  numerical calculations.) \textit{(Right)} Corresponding potential well
  for magnons and the first few levels in the well.}
\label{twexamp} 
\end{figure}

One to determine the parameter $\lambda$ is to measure
how the texture in a sample of $^3$He-B changes with the density of vortex
lines while all other contributions to the textural energy are kept
constant. In practice this requires stable temperature and pressure and
the absence of a global counterflow, i.e. measurements in the equilibrium vortex
state as a function of the angular velocity~$\Omega$.

Our sample is in a cylindrical container with a diameter of 6\,mm. The axis of the
cylinder is parallel to both magnetic field and rotation velocity with
a precision of about 1$^\circ$. In such a sample the texture is
usually of the axially symmetric flare-out type: The orbital anisotropy
vector $\hat{\bf l}$ is oriented along the axis in the center of the sample
and smoothly reorients perpendicular to the walls of the container at the
sample edges. The polar angle $\beta_l$ of the $\hat{\bf l}$-vector grows
approximately linear at small radii $r$ as shown in Fig.~\ref{twexamp}
(left). Since the effect from vortices is to turn vector $\hat{\bf l}$ away
from the axial orientation, with increasing $\lambda$ the slope
$\beta'_l = d\beta_l/dr|_{r=0}$ increases. It is this slope which we probe in
the experiment.

The remarkable feature of superfluid $^3$He, which makes NMR a useful
tool for probing the order parameter, is the spin-orbit interaction in the
Cooper pairs. In our case of the B phase and small tipping angles $\beta_M$ of
the magnetization the energy of spin-orbit interaction is \cite{qballprl}
\begin{equation}
F_{\rm so} = 
4\frac{\chi}{\gamma^2}\Omega_B^2\left
(\frac25 \sin^2\frac{\beta_l}2 \sin^2\frac{\beta_M}2 - 
\sin^4\frac{\beta_l}2 \sin^4\frac{\beta_M}2\right),
\label{fso}
\end{equation}
where $\chi$ is the magnetic susceptibility, $\gamma$ is the gyromagnetic ratio
and $\Omega_B$ is the Leggett frequency in the B phase. At small radii (and
thus small $\beta_l$) the fourth-order term in the expression for $F_{\rm
  so}$ can be neglected. Additionally $\sin^2(\beta_l/2) \approx
(\beta_l/2)^2 \propto (\beta'_l r)^2$, and thus $F_{\rm so}$ has the shape of
a harmonic potential well close to the axis of the sample. In alternative
language one can say that the texture forms a nearly-harmonic trap for
magnon excitations.

\begin{figure}
\centerline{\includegraphics[scale=1]{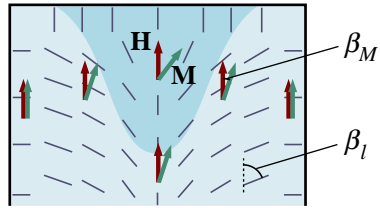}}
\caption{(color online) Schematic representation of a coherently precessing
  state which corresponds to magnon condensation at an energy level of the
  trap, formed by the order-parameter texture. The thin bars represent
 the  direction of the orbital anisotropy vector $\hat{\bf l}$.}
\label{pssketch} 
\end{figure}

An analysis of the magnetization dynamics in the presence of such a potential
\cite{vwspinwaves} leads to the conclusion that there exist localized spin
waves with frequencies $\omega_{\rm n}$ which correspond to levels in the
textural potential well:
\begin{equation}
\omega_{\rm n} - \omega_{\rm L} = \sqrt{\frac{96}{325}} 
\frac{\Omega_B^2}{\omega_{\rm L}} \xi_D \beta'_l ({\rm n} + 1),
\label{lev1}
\end{equation}
where $\omega_{\rm L} = |\gamma| H$ is the Larmor frequency, $\xi_D$ is the
dipolar length and n is the quantum number.
In the non-linear regime (when magnon filling of
a level is sufficiently high) such a spin wave can also be called a Bose-Einstein condensate of
magnons in a trap \cite{20ybec}. In this paper we use both languages interchangeably.
Some similarities of such precessing states to the Q-balls of high energy
physics were also suggested~\cite{qballprl}.

The frequencies of the trapped spin waves can be determined with NMR as
described in the following section. As shown in Fig.~\ref{twexamp} (right),
these frequencies are a sensitive tool to probe the texture and thus the value
of the vortex textural energy parameter $\lambda$. Note that for the spatially
homogeneous rf field in a typical NMR experiment coupling exists only with
spin wave states of even quantum number n.

In our experiment the texture is actually not axially uniform, since the
NMR pick-up coil is placed close to the top end plate of the sample
cylinder. Moreover the inhomogeneity of the static NMR field can
provide additional axial trapping in a local minimum of the field due to
the Zeeman energy. As a result, the trap for magnons is modified and becomes
3-dimensional, as schematically shown in
Fig.~\ref{pssketch}. Correspondingly the levels for the spin waves become
\begin{equation}
\omega_{\rm nm} = \omega_{\rm L} + \omega_r ({\rm n} + 1) +
      \omega_z ({\rm m} + 1/2).
\label{lev2}
\end{equation}
Here m is a new axial quantum number and the trapping frequency $\omega_z$
has some fixed value which is not under experimental control in this measurement. The
radial trapping frequency $\omega_r$ is determined by the texture as before in
Eq.~(\ref{lev1}). In our experiment $\omega_z \ll \omega_r$.

Note that the precessing states discussed here represent coherent
precession of the magnetization with the same phase over the whole volume,
despite magnetic field and textural inhomogeneities. The amplitude of the
precession $\beta_M$, however, changes over the volume as prescribed by the
wave function $\psi$ of the magnon condensate in the potential well since
$\sin^2(\beta_M/2) \propto (S-S_z)/\hbar$ is proportional to the density of
magnons $|\psi|^2$.

\section{CW NMR response}
\label{sec:nmr}

\begin{figure}
\centerline{\includegraphics[scale=0.9]{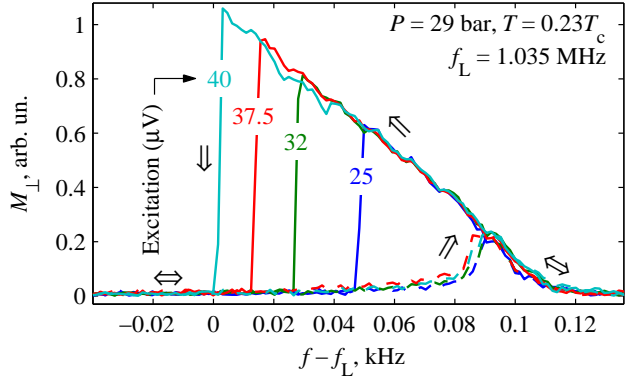}}
\caption{(color online) CW NMR response of the ground level in the textural
  potential well in stationary ($\Omega=0$) $^3$He-B for different
  excitation amplitudes. Total
  transverse magnetization $M_\perp$ is plotted against the frequency shift
  from the Larmor frequency $f_{\rm L}$. Arrows indicate the sweep
  direction.}
\label{psexc} 
\end{figure}

The $^3$He-B sample is installed in the rotating nuclear demagnetization
cryostat, where it can be cooled to below $0.2\,T_{\rm c}$ in rotation with velocities up to
3.5\,rad/s.  The measurements have been done at the pressure $P = 29\,$bar
at two values of the static magnetic NMR field which correspond to the
Larmor frequencies $f_{\rm L} = \omega_{\rm L}/(2\pi) = 1.035\,$MHz and
0.894\,MHz. The NMR pick-up coil is part of a tuned tank circuit with a Q
value above 5000. In the cw NMR measurement the rf excitation frequency $f$
is kept fixed at the resonance frequency of the tank circuit. The NMR
spectra are measured as a function of the frequency shift $f-f_{\rm L}$ by
sweeping the magnetic field and thus the Larmor frequency $f_{\rm L}$.
The temperature is measured from the damping of a quartz tuning fork
oscillator~\cite{fork}, which is calibrated against a $^3$He melting-curve
thermometer. More details on the experimental setup can be found in
Ref.~\cite{expsetup}.

When the ground level in the textural potential well (corresponding to
${\rm n} = {\rm m} = 0$ in Eq.~(\ref{lev2})) is crossed during a downward
sweep of frequency (upward sweep of magnetic field), the magnon condensate
starts to build up, Fig.~\ref{psexc}. The total transverse magnetization
$M_\perp = M \langle\sin\beta_M\rangle$ is related to the number of magnons
in the condensate $N_{\rm magn}$. For small $\beta_M$ (which is not always
the case) $N_{\rm magn} \propto M_\perp^2$. When the number of magnons in
the condensate grows, the resonance frequency decreases due to
magnon-magnon interactions. (This dependence is not included in
Eq.~(\ref{lev2}).) The interaction comes from two effects: First, the
negative fourth-order term in the spin-orbit interaction energy
(\ref{fso}). Second, the texture is not rigid enough and yields under the
influence from the precessing spin (owing to the spin-orbit
interaction): It becomes flatter in the part of the sample close to the
axis and $\beta'_l$ decreases \cite{qballprl}. 

The dependence of the
precession frequency on the number of magnons is a characteristic of a
particular state. That is why $M_\perp$ in Fig.~\ref{psexc} does not depend
on the amplitude of the rf excitation. This non-linearity is the reason why
such a NMR state is created
only in one direction of the NMR field (or frequency) sweep. To find the frequency
of the ground level in the textural potential well which corresponds to
the absence of magnons we fit a straight line to the lower part of the $M_\perp$ vs
$f-f_{\rm L}$ dependence and take its intersection with the baseline.

\begin{figure}
\centerline{\includegraphics[scale=1]{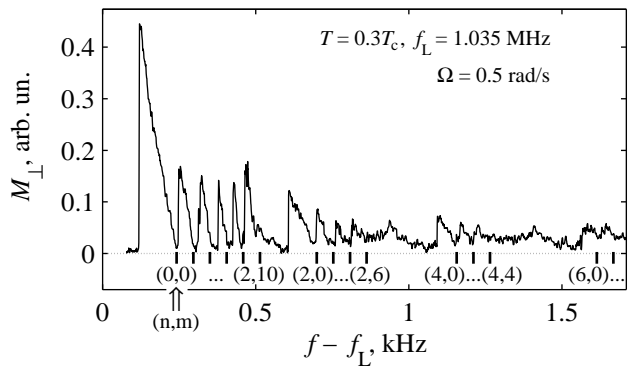}}
\caption{Series of resonances with
  harmonic oscillator spacing in the cw NMR spectrum of rotating
  vortex-free $^3$He-B,
  enumerated using radial (n) and axial (m) quantum numbers in
  a 3-dimensional potential well formed by the texture. }
\label{psspec} 
\end{figure}

Other levels in the magnon trap can be identified in a similar way, as
shown in Fig.~\ref{psspec}. This example is measured in vortex-free
rotation. The groups of peaks corresponding to different radial quantum
numbers n are clearly visible. The closely spaced peaks within each group
correspond to different axial quantum numbers m. The vertical lines below
the $M_\perp = 0$ axis represent fits to Eq.~(\ref{lev2}) with
$\omega_r/2\pi=0.23\,$kHz and $\omega_z/2\pi=27\,$Hz. The plot shows that
the trap is indeed nearly harmonic close to the bottom of the potential well.

By changing the rotation velocity one can verify that $\omega_r$ varies as
expected owing to the texture changes while $\omega_z$ remains constant
\cite{excbec}. Later in this paper we are only interested in the dependence of
$\omega_r(\Omega)$ on the rotation velocity, or more precisely, only in the shift with
respect to its value $\omega_r(0)$ without rotation. As can be seen from
Eq.~(\ref{lev2}) this shift coincides with the shift of the ground-level
peak $\omega_{00}(\Omega) - \omega_{00}(0)$, which is the quantity measured in
the experiment.

\section{Determination of the core contribution}
\label{sec:res}

\begin{figure}
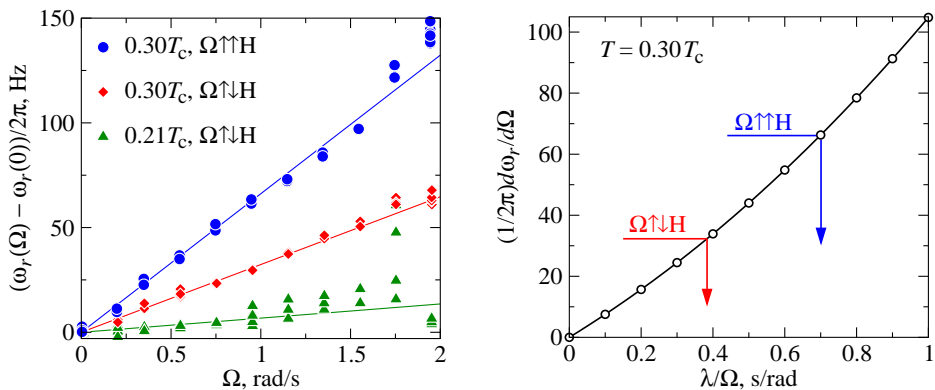

\includegraphics[scale=0.45]{psorig_fit.eps} \hfill
\includegraphics[scale=0.45]{lambda_examp.eps}
\caption{(color online) Measurements of the core contribution.
  \textit{(Left)} In the equilibrium vortex state the shift of the ground level of the
  magnon condensate with respect to its position at $\Omega = 0$ shows
  linear dependence on the angular velocity at $\Omega < 1\,$ rad/s.
  \textit{(Right)} Comparison of the slope $d\omega_r/d\Omega$, obtained
  from the experiment (horizontal lines) with that obtained from numerical
  calculations of the order-parameter texture for different values of
  $\lambda/\Omega$ (circles) yields the measured value of $\lambda/\Omega$
  (vertical arrows).}
\label{psorig} 
\end{figure}

The measurement proceeds as follows: At a constant temperature below
$0.4\,T_{\rm c}$ the rotation velocity is increased to more than 2\,rad/s. This
leads to vortex formation \cite{slowform} and ultimately to the
equilibrium vortex state. After that rotation is reduced in steps
while the ground level spin wave resonance is continuously monitored. At
any
given rotation velocity we wait until the frequency of the level stops changing.
After initial spin-up this may take more than an hour (see
Sec.~\ref{sec:dyn}), but after smaller downward steps in $\Omega$
the relaxation time is shorter. Finally we plot the shift of the ground-level
resonance with respect to its position at $\Omega=0$ versus $\Omega$, see
Fig.~\ref{psorig} (left) for examples.

The shift is due to the vortex contribution to the textural energy. Two
examples at $0.3\,T_{\rm c}$ for vortices which are parallel or
antiparallel to the magnetic field direction show that such vortices
exert clearly different effect on the texture, which proves the
gyromagnetic effect. The third example at $0.21\,T_{\rm c}$
demonstrates that at low temperatures the effect from the vortices
antiparallel to the magnetic field is small but still measurable.

As can be seen from the figure, this shift is a linear function of $\Omega$
at rotation velocities below 1\,rad/s. The slope
$d\omega_r(\Omega)/d\Omega$ is converted to the corresponding value of
$\lambda/\Omega$ in the following manner. For the conditions of the
experiment (pressure, temperature, magnetic field) we numerically calculate
the texture on a grid of $\lambda/\Omega$ and $\Omega$ values. We use the
calculation method and the $^3$He-B parameters from Ref.~\cite{juhaprog}. For each
texture we determine the $\omega_r$ value using the slope $\beta'_l$ at $r=0$ and
Eq.~(\ref{lev1}). As in the experiment, $\omega_r$ turns out to be a linear
function of $\Omega$ at $\Omega < 1\,$rad/s. The slope $d\omega_r/d\Omega$,
as determined from numerical calculations for $T=0.3\,T_{\rm c}$, is plotted
in Fig.~\ref{psorig} (right) with circles. As one can see, this dependence
is monotonic and smooth. We then use interpolation to convert
the experimentally measured slopes $d\omega_r/d\Omega$ to the corresponding values
of $\lambda/\Omega$.

\begin{figure}
\centerline{\includegraphics[width=0.65\linewidth]{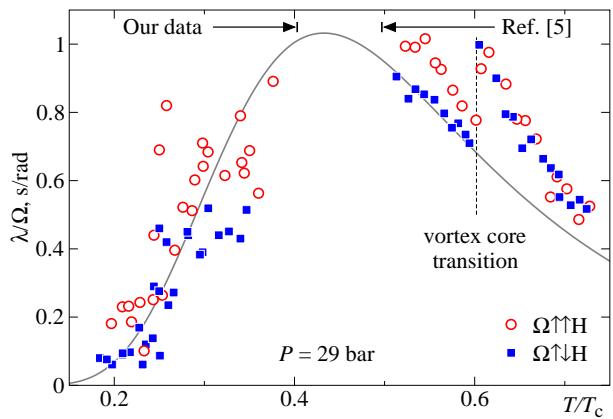}}
\caption{(color online) Textural energy parameter $\lambda/\Omega$ as a
  function of temperature. Our data is at $T < 0.4T_{\rm c}$. At
  temperatures above $0.5\,T_{\rm c}$ the data from Ref.~\cite{HakonenJLTP}
  are plotted, adapted as described in the text. The discontinuity at
  $0.6\,T_{\rm c}$ signifies the first order transition in the vortex core
  structure. Vortices oriented along and opposite to the magnetic field
  direction have different textural energy contributions owing to the
  gyromagnetic effect. The solid line is a theoretical estimation from
  Ref.~\cite{Thuneberg} which includes only the effect from the counterflow
  circulating around vortex cores.}
\label{flambda} 
\end{figure}

All our results for $\lambda/\Omega$ for the two vortex orientations in the
temperature range $(0.19\div0.4)T_{\rm c}$ are collected in
Fig.~\ref{flambda}. Also included in the plot are data from
Ref.~\cite{HakonenJLTP} at the pressure 29.3 bar and $f_{\rm L} =
922.5\,$kHz.  These data were measured also with the spin-wave resonance
technique using comparison to texture calculations.  The texture
calculation techniques have significantly improved since publication of
Ref.~\cite{HakonenJLTP}. Therefore we reanalyzed the measured spin-wave
resonance frequencies from Ref.~\cite{HakonenJLTP} (Fig.~19a) using the
same texture calculation program as for our measurements. This resulted in
a slight upward correction for the values of $\lambda/\Omega$. These
corrected points are plotted at $T>0.5\,T_{\rm c}$. The solid line in the
plot shows the theoretical calculation of the flow-related contribution
$\lambda_{\rm f}/\Omega$ from Ref.~\cite{Thuneberg}.

In general the agreement is good between old measurements at higher
temperatures and our new
measurements at lower temperatures. The theoretical curve also
captures the overall trend and magnitude of $\lambda/\Omega$ well.
Our data points exhibit rather large scatter. One possible cause for this scatter
is the remaining disorder in the vortex cluster, when it is prepared at low
temperatures. As explained in the next section, when vortex lines are not
all perfectly aligned with the rotation axis, the effective value of $\lambda$
(determined by the average effect of all vortex lines) may change. At low
temperatures vortex formation proceeds typically through instabilities and interactions of
the existing vortices, often in a localized turbulent burst \cite{expsetup}. The
vortex configuration created in such a process is far from the equilibrium
one and its relaxation to the equilibrium state is impeded by the existence of many
local energy minima in the vortex configurations. In addition the small value of the
dissipative mutual friction parameter at low temperatures slows down the relaxation. One way to avoid
such problems would be to prepare the vortex cluster at higher temperatures
(above $0.7\,T_{\rm c}$) and cool it down in rotation to low
temperatures before doing the measurement. Such measurements would require 
more time. Additionally in this way there is a chance to miss a
transition in vortex core structure, if it is of first order and can be
substantially supercooled.

\begin{figure}
\centerline{\includegraphics[width=0.55\linewidth]{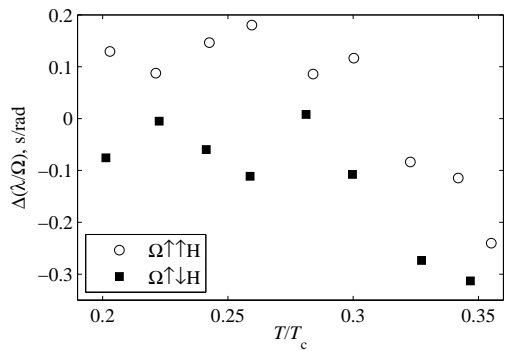}}
\caption{Difference between measured values of $\lambda/\Omega$ and
  theoretical estimation~\cite{Thuneberg} plotted as a function of
  temperature. The experimental points have been averaged in temperature
  bins of $0.02T_{\rm c}$ width. }
\label{lamdiff} 
\end{figure}

The flow-related contribution to $\lambda$ drops exponentially to zero as
the temperature decreases since the superfluid density anisotropy vanishes at
$T=0$. Thus the finite value of $\lambda$ in the $T\rightarrow0$ limit
signifies the contribution from the vortex cores. As can be seen from
Fig.~\ref{flambda}, indeed the measured values of $\lambda/\Omega$ approach
a constant in the $T\rightarrow0$ limit, which is different for the two
vortex orientations. To demonstrate this more clearly we plot in
Fig.~\ref{lamdiff} the experimental points averaged in narrow temperature
bins and additionally we subtract the theoretical estimation of the flow
contribution $\lambda_{\rm f}$ to $\lambda$. Indeed the resulting points are approximately
temperature independent below $0.3\,T_{\rm c}$. Unfortunately, the
theoretical calculations of the flow contribution are not precise enough to
determine the absolute value of the pure core contribution to $\lambda$.
The best way to achieve this goal will be to extend measurements to
temperatures of about $0.15\,T_{\rm c}$ where the flow contribution can be
completely ignored.

The magnitude of the gyromagnetic effect can be found by comparing
$\lambda$ for the two vortex orientations. The average difference between
$\lambda/\Omega$ for these orientations, i.e. $2\kappa/(\Omega H)$, is
found to be $0.18\pm0.07\,$s/rad. This difference should increase with
decreasing magnetic field inversely proportional to $H$. Thus
we can conclude that at about twice smaller magnetic field than used in the
present experiment, the value of $\lambda$ for $\Omega\uparrow\downarrow H$
orientation will become negative in $T\rightarrow 0$ limit. In fact, the
measurements in Fig.~\ref{flambda} are done for two slightly different
magnetic field magnitudes, but within the scatter the field dependence of
$\lambda$ is not observed.

An interesting question, which measurements of $\lambda$ can answer, is
whether on the way to zero temperature there are more transitions in the
vortex core structure. The core of the low-temperature vortex still
possesses 2-fold symmetry which in principle can be broken in other phase
transitions. However, in our measurements we have not found clearly
identifiable jumps in the temperature dependence of $\lambda$ with an
associated hysteresis which would indicate possible transitions in the
vortex core structure.

Finally we note that the core contribution to $\lambda$ should in fact
have some temperature dependence in the $T\rightarrow 0$ limit owing to
the existence of bound fermion states in the vortex core \cite{corefermions}.
The effect of the core-bound quasiparticles on $\lambda$ has not been
considered theoretically so far. From general principles this contribution
should have a power-law temperature dependence (most probably $\propto T^2$
\cite{volpriv}), since even the lowest
experimental temperatures are much larger than the so-called minigap,
the spacing between levels of the bound fermion states. This is unlike the situation in
the bulk, where $T\ll \Delta$ and the flow contribution to $\lambda$
freezes out fast, as $\exp(-\Delta/T)$. Thus the contributions from the
core-bound quasiparticles is expected to dominate the temperature
dependence of $\lambda$ in the $T\rightarrow 0$ limit. To observe it, though,
one has to improve the accuracy of measurements from that in
Fig.~\ref{lamdiff}. One possible improvement is to repeat the measurements at zero pressure. Since the size of the vortex core
increases towards lower pressure, the magnitude of the vortex core
contribution to textural energy should increase.

\section{Probing vortex dynamics}
\label{sec:dyn}

The techniques developed for the measurement of $\lambda$ can also be used
to study vortex dynamics. In typical measurements, which use linear NMR
response, one can track the development of the global azimuthal flow in the
sample using the so called counterflow peak in the NMR spectrum
\cite{expsetup,spinupdn}. The global axial flow produced by the twisted
vortex state affects the shape of the Larmor peak in the cw NMR spectrum
\cite{twisted}. Such techniques relies on the measurement of the
order-parameter texture, which depends on the distribution of the global
flow and vortices in the sample. 
The measurement of the texture using
spin-wave resonances, as performed in this work, is a similar technique.
However, its high sensitivity to the texture profile close to the axis of
the sample makes it especially useful for the cases when the global flow is
absent. Thus it can provide additional insight to the processes of
establishing equilibrium and order in the vortex cluster after a
perturbation, or to the loss of polarization and even
turbulence when time-dependent rotation drive is applied.

\begin{figure}
\centerline{\includegraphics[width=0.85\linewidth]{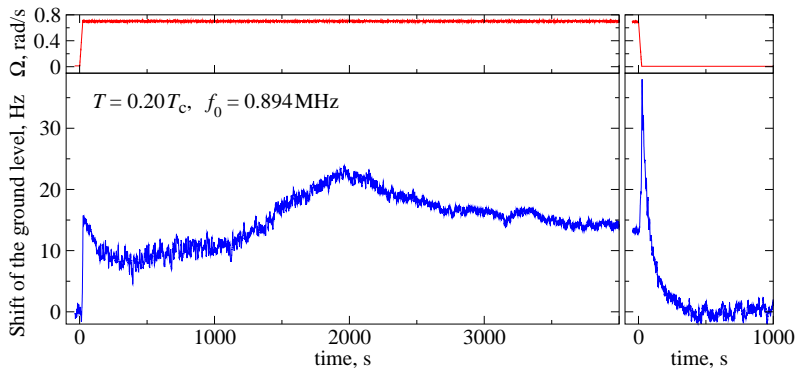}}
\caption{(color online) Spin-up \textit{(left)} and spin-down
  \textit{(right)} of the $^3$He-B sample probed
  by monitoring the ground magnon level in the textural potential well. Plots
in the top row show the angular velocity. In the bottom row the shift of the
ground level with respect to its position at $\Omega=0$ is shown. Here the
shift is determined by continuously monitoring the amplitude of the condensate
signal at a fixed frequency shift. The change in the amplitude is recalculated
to frequency shifts using the measured slope of amplitude vs. frequency
dependence as in Fig.~\ref{psexc}. Note that the time axes scaling is the
same for spin-up and spin-down and thus relaxation after spin-up proceeds
much longer than after spin-down. For a discussion of the various features seen
in the plots see the text.}
\label{spinupdn}
\end{figure}

When vortex lines are not completely polarized along the rotation axis, their
effect on the texture is renormalized and the parameter $\lambda$ in the
textural energy is replaced by some effective parameter $\lambda_{\rm eff}$
which depends on the particular vortex configuration. In Ref.~\cite{juhalt}
the model of the equilibrium vortex cluster with superimposed fluctuations
is considered. If the vorticity $\bm{\omega}_{\rm v}$ is decomposed as $\bm{\omega}_{\rm v} =
\bm{\Omega} + \bm{\omega}_{\rm v}'$ in an equilibrium part $\bm{\Omega}$ and
a random part $\bm{\omega}_{\rm v}'$ with $\langle\bm{\omega}_{\rm
  v}'\rangle = 0$, then
\begin{equation}
\lambda_{\rm eff} = \lambda \frac{1+(\omega_{{\rm v}\parallel}/\Omega)^2
     - (\omega_{{\rm v}\perp}/\Omega)^2}{\sqrt{\strut 1+
       (\omega_{{\rm v}\parallel}/\Omega)^2
     + 2(\omega_{{\rm v}\perp}/\Omega)^2}}~,
\label{leff}
\end{equation}
where $\omega_{{\rm v}\parallel}^2 = \langle \omega_{{\rm v}z}'^2\rangle$
and $\omega_{{\rm v}\perp}^2 = \langle \omega_{{\rm v}x}'^2\rangle =
\langle \omega_{{\rm v}y}'^2\rangle$. Here the coordinate axis $z$ is directed
along the
rotation axis and the random vorticity is assumed isotropic in the $xy$
plane perpendicular to rotation axis. As
can be seen from Eq.~(\ref{leff}), the value of $\lambda_{\rm eff}$ can be
larger than $\lambda$, if $\omega_{{\rm v}\parallel}$ dominates, or it can be
smaller than $\lambda$, or even of different sign than $\lambda$, if
$\omega_{{\rm v}\perp}$ dominates.

The time dependence of $\lambda_{\rm eff}$ can be observed through the frequency shift of the ground
magnon level in the textural potential well. In Fig.~\ref{spinupdn} we plot
the responses to a rapid increase of the rotation velocity from rest
(so-called spin-up) and to a rapid stop of rotation from the equilibrium vortex state (spin-down). These processes were studied previously at low
temperatures from the linear NMR response of the
counterflow peak \cite{spinupdn}. The conclusion from these measurements
was that the spin-down process is always laminar in the bulk volume, with a
smoothly decreasing density of
almost straight vortex lines. The spin-down response
in Fig.~\ref{spinupdn} (right) agrees with this conclusion: Here the ground
level resonance first moves to higher frequencies due to the build up of
counterflow between the stationary normal component and superfluid component,
which still rotates due to the presence of vortices. The relaxation of the
vortex density to zero happens within a time which is in agreement with the
measurements of
Ref.~\cite{spinupdn}.

\begin{figure}
\centerline{\includegraphics[width=0.8\linewidth]{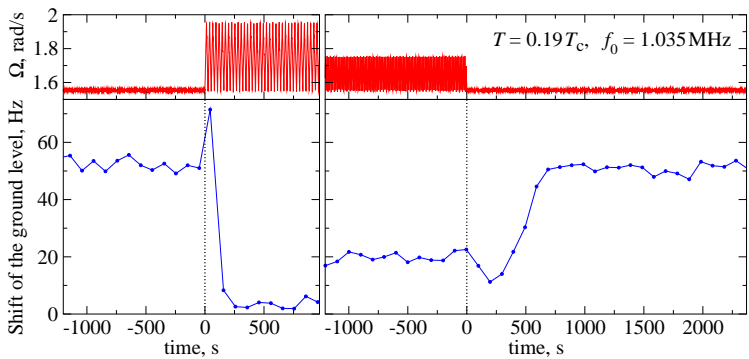}}
\caption{(color online) Vortex lines in rotation with oscillating component
  in the rotation drive. Plots in the top row show the angular velocity. In
  the bottom row the shift of the ground level of the magnon condensate
  with respect to its position at $\Omega = 0$ is plotted. \textit{(Left)}
  Here linear modulation of $\Omega$ between 1.55 and 1.95\,rad/s with the
  rate 0.03\,rad/s$^2$ is switched on at $t=0$. Vortex lines seem to
  significantly depolarize within 200\,s and the effective $\lambda$ drops
  by more than an order of magnitude. \textit{(Right)} A smaller modulation
  of $\Omega$ between 1.55 and 1.75\,rad/s with the same rate
  0.03\,rad/s$^2$ results in a smaller loss of polarization: Here the
  effective $\lambda$ is almost half of the equilibrium value. The
  transient process after switching off the modulation of $\Omega$ proceeds
  slower than the transient when the modulation is switched on.}
\label{osc} 
\end{figure}

The response to spin-up in Fig.~\ref{spinupdn} (left) shows more features.
Previous measurements \cite{slowform,spinupdn} demonstrated
that the spin-up behaviour depends on the number and the
configuration of the pre-existing vortices (at zero rotation), so-called
dynamic remnants \cite{remnants}, which act as seeds for the creation of new
vortices. For the spin-up in Fig.~\ref{spinupdn} the waiting time at $\Omega = 0$ after the previous spin-down was
25\,min and estimated number of remnants is of the order of hundred.
The initial shift of the resonance peak to higher frequencies after the
increase of $\Omega$ is explained by the increase of the azimuthal
counterflow velocity. As new vortices are formed the counterflow velocity and the frequency shift of the magnon level
decrease. Eventually
the vortex contribution to the textural energy becomes significant and the
peak again moves to higher frequency. In a smooth laminar process the evolution of the frequency shift
of the ground magnon level would have only one minimum between the values
dominated by the counterflow and vortex cluster. The dependence in
Fig.~\ref{spinupdn} has one extra maximum (at time around 2000\,s). Thus the simple model of laminar spin-up is not applicable here and contribution from turbulent bursts, which can build $\omega_{{\rm v}\parallel}$ component in Eq.~(\ref{leff}), is possible.

An extra vorticity in the horizontal plane, perpendicular to rotation
velocity, leads to a decrease of the effective
value of $\lambda$. Such vorticity appears, for example, in Kelvin-wave
excitations of individual vortex lines or global oscillating modes of the
vortex array \cite{bar}. One can try to excite such modes with oscillating
$\Omega$ drives. Two examples of measurements are shown in
Fig.~\ref{osc}. As can be seen from these plots, increasing the amplitude
of the oscillating component in the drive indeed leads to a decreasing
frequency shift of the ground state. Possible interpretation of this effect
is that the effective value of
$\lambda$ decreases. One can make $\lambda_{\rm eff}$ to be almost zero which would
require the vortex line length in the horizontal plane to be comparable with
the line length along the rotation axis. Unfortunately it is impossible to
determine the exact vortex configuration from the measurement of the single
parameter $\lambda_{\rm eff}$. More detailed studies of the transients, when
the modulation is switched on and off, can in principle provide additional
information.

\section{Conclusions}

The spin-wave resonance peaks in the cw NMR absorption spectrum
of $^3$He-B provide sensitive probes of the order-parameter texture and
useful tools to determine textural parameters. Previously they were used
at temperatures above $0.5\,T_{\rm c}$ to measure, for example, the dipolar
coherence length $\xi_D$ and the textural energy parameter
$\lambda$ for vortices. Recently a non-linear modification of such spin waves or, in
alternative language, a Bose-Einstein condensate of magnons in a trap,
created by the order-parameter texture, was discovered at low temperatures.
We used this novel precessing state to extend the measurement of $\lambda$
from $0.4\,T_{\rm c}$ to below $0.2\,T_{\rm c}$. Earlier information on
$\lambda$ in this temperature range was only available from complicated and
less reliable analysis of the overall shape of the linear NMR absorption
spectrum.

We found that the main temperature dependence of $\lambda$ in this
temperature range comes from the flow circulating around vortex cores, in
agreement with theoretical calculations. This contribution decreases
exponentially to zero, when the temperature approaches zero. Thus from the
measurements at the lowest temperatures the vortex core contribution to
$\lambda$ can be determined. The gyromagnetic effect, i.e. the difference
between the values of $\lambda$ for vortices parallel and antiparallel to
the magnetic field, is clearly observed in the whole temperature range
studied. No sign of another transition in the vortex core structure was found (in
addition to the axial symmetry breaking at $0.6\,T_{\rm c}$).

Our preliminary measurements indicate that the magnon BEC can be a useful tool
to probe vortex dynamics, in particular wave excitations on vortex lines.

\begin{acknowledgements}
  We thank Yu.M. Bunkov and G.E. Volovik for stimulating discussions and
  R.E. Solntsev for help with the experiment. This work was partially
  supported by the Academy of Finland and by the EU-funded Integrating
  Activity project MICROKELVIN.
\end{acknowledgements}


\begin{thebibliography}{}
% etc

\bibitem{vortrev} M.M. Salomaa and G.E. Volovik,
  Rev. Mod. Phys. \textbf{59}, 533 (1987).

\bibitem{coretrans} O.T. Ikkala, G.E. Volovik, P.J. Hakonen, Yu.M.
  Bunkov, S.T. Islander, and G.A. Kharadze, Pis'ma Zh. Eksp.
    Teor. Fiz. {\bf 35}, 338 (1982) [JETP Lett. {\bf 35}, 416
  (1982)].

\bibitem{Thuneberg} E.V. Thuneberg, J. Low Temp. Phys. \textbf{122}, 657
  (2001).

\bibitem{gyro} P.J. Hakonen, M. Krusius, M.M. Salomaa, J.T. Simola, Yu.M.
  Bunkov, V.P. Mineev and G. E. Volovik, Phys. Rev. Lett. \textbf{51}, 1362
  (1983).

\bibitem{HakonenJLTP} P.J. Hakonen, M. Krusius, M.M. Salomaa, R.H.
  Salmelin, J.T. Simola, A.D. Gongadze, G.E. Vachnadze, and G.A. Kharadze,
  J. Low Temp. Phys.  \textbf{76}, 225 (1989).

\bibitem{qballprl} Yu.M. Bunkov and G.E. Volovik, Phys. Rev. Lett.
  \textbf{98}, 265302 (2007).

\bibitem{vwspinwaves} D. Vollhardt and P. W\"olfle, \textit{The Superfluid
    Phases of Helium 3}, pp. 387--388. Taylor \& Francis, London (1990).

\bibitem{20ybec} G.E. Volovik, J. Low Temp. Phys. \textbf{153}, 266 (2008).

\bibitem{fork} M. Bla\v{z}kov\'a, M. \v{C}love\v{c}ko, V.B. Eltsov,
  E. Ga\v{z}o, R. de Graaf, J.J. Hosio, M. Krusius, D. Schmoranzer,
  W. Schoepe, L. Skrbek, P. Skyba, R.E. Solntsev, and W.F. Vinen, J. Low
  Temp. Phys. \textbf{150}, 525 (2008).

\bibitem{expsetup} R. de Graaf, R. H\"anninen, T.V. Chagovets, V.B. Eltsov, M. Krusius,
  and R.E. Solntsev, J. Low Temp. Phys. \textbf{153}, 197 (2008).

\bibitem{excbec} Yu.M. Bunkov, V.B. Eltsov, R. de Graaf, P.J. Heikkinen,
  J.J. Hosio, M. Krusius, and G.E. Volovik, arXiv:1002.1674.

\bibitem{slowform} A.P. Finne, V.B. Eltsov, G. Eska, R. H\"anninen, J.
  Kopu, M.  Krusius, E.V. Thuneberg, and M. Tsubota, Phys. Rev. Lett.
  \textbf{96}, 085301 (2006).

\bibitem{juhaprog} J. Kopu, J. Low Temp. Phys. \textbf{146}, 47 (2007).

\bibitem{corefermions} N.B. Kopnin, and G.E. Volovik, Phys. Rev. B \textbf{57},
  8526 (1998).

\bibitem{volpriv} G.E. Volovik, private communication.

\bibitem{spinupdn} V.B. Eltsov, R. de Graaf, P.J. Heikkinen, J.J. Hosio,
  R. H\"anninen, M. Krusius, and V.S. L'vov, arXiv:1005.0546.

\bibitem{twisted} V.B. Eltsov, A.P. Finne, R. H\"anninen, J. Kopu, M.
  Krusius, M. Tsubota, and E.V. Thuneberg, Phys. Rev. Lett. \textbf{96},
  215302 (2006).

\bibitem{juhalt} J. Kopu, V.B. Eltsov, A.P. Finne, M. Krusius, and
  G.E. Volovik, AIP Conference Proceedings \textbf{850}, 181 (2006).

\bibitem{remnants} R.E. Solntsev, R. de Graaf, V.B. Eltsov, R. H\"anninen, and
  M. Krusius, J. Low Temp. Phys. \textbf{148}, 311 (2007).

\bibitem{bar} K.L. Henderson and C.F. Barenghi,
  Europhys. Lett. \textbf{67}, 56 (2004).

%\bibitem{coretransdiagr}  J.P. Pekola, J.T. Simola, P.J. Hakonen,
%  M. Krusius, O.V. Lounasmaa, K.K. Nummila, G. Mamniashvili, R. E. Packard,
%  and G.E. Volovik, Phys. Rev. Lett. \textbf{53}, 584 (1984).

\bibitem{coretransdiagr} M. Krusius, P.J. Hakonen, and J.T. Simola, Physica
  \textbf{126B}, 22 (1984).

\bibitem{corecalc} E.V. Thuneberg, Phys. Rev. B \textbf{36}, 3853 (1987).

\end{thebibliography}


\end{document}